\documentclass[conference]{IEEEtran}
\IEEEoverridecommandlockouts
\usepackage{cite}
\usepackage{amsmath,amssymb,amsfonts}
\usepackage{algorithmic}
\usepackage{graphicx}
\graphicspath{ {./figures/} }
\usepackage{textcomp}
\usepackage{xcolor}
\usepackage{lscape}
\usepackage{multirow}
\usepackage[euler]{textgreek}
\usepackage{tabulary}
\usepackage{amsmath}
\usepackage{balance}

\usepackage{booktabs}
\newcommand{\tabitem}{~~\llap{\textbullet}~~}

\def\BibTeX{{\rm B\kern-.05em{\sc i\kern-.025em b}\kern-.08em
    T\kern-.1667em\lower.7ex\hbox{E}\kern-.125emX}}
\begin{document}

\makeatletter
\newcommand{\linebreakand}{%
  \end{@IEEEauthorhalign}
  \hfill\mbox{}\par
  \mbox{}\hfill\begin{@IEEEauthorhalign}
}
\makeatother

\title{Exploratory Analysis of a Social Media Network in Sri Lanka during the COVID-19 Virus Outbreak}

\author{
\IEEEauthorblockN{Damitha Lenadora}
\IEEEauthorblockA{\textit{Computer Science and Engineering} \\
\textit{University of Moratuwa}\\
Katubedda, Sri Lanka \\
damitha.15@cse.mrt.ac.lk}
\and
\IEEEauthorblockN{Gihan Gamage}
\IEEEauthorblockA{\textit{Computer Science and Engineering} \\
\textit{University of Moratuwa}\\
Katubedda, Sri Lanka \\
gihangamage.15@cse.mrt.ac.lk}
\and
\IEEEauthorblockN{Dilantha Haputhanthri}
\IEEEauthorblockA{\textit{Computer Science and Engineering} \\
\textit{University of Moratuwa}\\
Katubedda, Sri Lanka \\
dilantha.15@cse.mrt.ac.lk}
\linebreakand
\IEEEauthorblockN{Dulani Meedeniya}
\IEEEauthorblockA{\textit{Computer Science and Engineering} \\
\textit{University of Moratuwa}\\
Katubedda, Sri Lanka \\
dulanim@cse.mrt.ac.lk}
\and
\IEEEauthorblockN{Indika Perera}
\IEEEauthorblockA{\textit{Computer Science and Engineering} \\
\textit{University of Moratuwa}\\
Katubedda, Sri Lanka \\
indika@cse.mrt.ac.lk}
}

\maketitle

\begin{abstract}

During the COVID-19 pandemic, multiple aspects of human life  were subjected to unprecedented changes, globally. In Sri Lanka, a developing country located in South Asia, it was possible to observe a range of events that arose due to the influence of the COVID-19 virus outbreak. Thus, the people of Sri Lanka used Social Media to voice their opinions regarding such events and those involved in them, enabling the ideal avenue to explore the social perception. However, the outcome of such actions was at certain times detrimental. This study was conducted as an attempt to identify the reasons for such instances as well as to identify the behaviours of the Sri Lankan populace during such a crisis event. To support this study, observations, as well as data of related posts from a sample of 50 sources, were manually collected from the most popular social media platform in Sri Lanka, Facebook. The posts considered spanned until approximately a month after the initial major virus outbreak in the country and contained content that even vaguely related to the virus. Utilising such data, various forms of analyses such as topic significance and topic co-occurrences were conducted. The findings highlight, while there can be social detrimental ideas shared, the majority of the posts point constructive and positive thoughts suggesting the successful influence from the cultural and social values Sri Lanka society promotes throughout.

\end{abstract}


\section{Introduction}
The World Health Organisation (WHO) defined the COVID-19 outbreak as a global pandemic and a severe global threat.
Its effects spanned to nearly every aspect of human life around the globe, including health, economy, environment as well as society. 
In the attempts of managing such outbreaks, information plays a vital role, because of its ability to enhance or even hinder the countermeasures deployed by authorities \cite{Kim2019}.
For example, in Sri Lanka, a developing country with a population of approximately 21 million \cite{populationSL}, social media anticipated a possible country-wide lock-down before it was officially declared by the government. 
As a result, people overcrowded shops and markets to buy essential goods without following any health precautions. 
This also led to a shortage of certain resources, especially for those who got through life by a daily wage and could not stockpile any essential goods.
These actions hindered the government which was attempting to control and mitigate the effects of the COVID-19 virus, and potentially increased the vulnerability of the society against the pandemic. 
In addition, another notable incident that occurred was, the spreading severe misinformation on social media during the outbreak. As a response to such actions, the government took measures to incarcerate the individuals involved. 
When facing an pandemic like COVID-19 for which there were no cures nor vaccines during the first few months, the only tools that humans held were social distancing and quarantine.
Thus, the value of researching and subsequently harnessing social media to improve countermeasures against the virus in the prevailing information era is apparent.

Social media platforms such as Facebook provide direct access to an immense amount of information, which also includes fact-based news as well as rumours. 
At the same time, these platforms encourage various perceptions based on information provided by the society. In addition, the conception of related narratives also occur. 
This sparks and influences public debates, especially when the topics discussed are controversial. 
Furthermore, it is observed that users of social media tend to acquire or ignore information based on such perceptions \cite{cinelli2020covid19:italypaper}.

Thus, with the objective to understand the behaviours of the Sri Lankan people during a pandemic, this research provides an analysis based on observations as well as content posted by a sample set of the Sri Lankan Facebook user-base. The sample set includes data from posts of 50 randomly selected profiles and pages that span across approximately 60 days. 
A notable point regarding this data collection process was that it was manually done only considering publicly shared content, which was subsequently anonymised. 
Despite its severe inefficiencies, this manual process was chosen due to it being one of the few non-restrictive forms of data collection on Facebook. 
Although this method imposed limitations on the number of sources that could practically be extracted, this allowed for a much more thorough analysis which considered not only image based posts, but also posts that indirectly related the pandemic.   
The results of the research present behavioural patterns observed in three major topic areas as well as analyses aggregating the collected data.
Utilising the results of this study, the possibility of using social media as a means of mitigating the spread of the pandemic by governments and relevant authorities is also present. In addition, it should be noted that the scenario chosen for this research may be generalisable to any developing country as well as any crisis event.

\section{Related work}

The impact that social media has on society is growing day by day. The field of research is also no exception \cite{Kapoor2018:SM}. The analysis of how the general populace reacted in social media during significant events is nothing new as well. 
Moody-Ramirez and Church \cite{MoodyRamirez2019} discusses how citizens of the United States of America (USA) have used memes in Facebook to convey their political opinions the during the 2016 USA presidential elections. The research by Sandoval-Almazan and Valle-Cruz \cite{SandovalAlmazan2018} presents the behaviour of Facebook users during a local government campaign of the Central State of Mexico in 2017. Here, authors have used more than 4000 Facebook posts to examine the people's sentiments to the campaign.

The study by Öztürk and Ayvaz \cite{OZTURK2018136} can be taken as a prime example where the significant event is a crisis. In this research, the authors have collected Tweets related to the Syrian refugee crisis and performed a sentiment analysis on them. Here a noteworthy point would be that the authors analysed English and Turkish Tweets separately. Whereupon doing as such, they have discovered a significant difference in terms of sentiments. This sheds insight into how peoples' reactions would differ based on their ethnicity. 
Additionally, several researches \cite{7857213, Li2015TwitterMF, PUROHIT20132438, CARLEY201648} present methods to utilise messages in social media to aid in the situation at hand.

Falling under a subsection of crisis events, infectious diseases also possess a significant threat to humanity as they possess the capability to hamper our everyday lives in more ways than one. Thus, social media content relating to such events are quite common during outbreaks of such diseases. This was especially the case during the Middle East Respiratory Syndrome (MERS) \cite{OMRANI2015188} virus outbreak in South Korea. 
The research by Kim et al. \cite{MERS:Korea1} utilises such content where the authors have analysed social media messages, especially considering their spatiotemporal attributes. Here the authors have proposed that the local governments must play an active role to supply official information as the excessive spread of rumours and misinformation was a significant issue during the crisis. Similarly, the study by Choi et al. \cite{article:MERSKorea} which focuses on the outbreak in South Korea as well, also signifies the impact of social media on risk perceptions. Although not limited Social Media alone, Jang and Baek \cite{MERSPHO} presented yet another research related to the MERS outbreak in South Korea. Here, the authors stressed the importance of the credibility of Public Health Officials in order to avoid the spread of misinformation.   
In addition, the reaction of the Chinese populace on social media due to the MERS-CoV
and Avian Influenza A(H7N9) outbreaks has been analysed by Fung et al. \cite{pmid24359669:MERS}, where they have concluded that the distance to a particular outbreak of the disease plays a key role on how strongly the people would react.

COVID-19 which is yet another infectious disease that has had a considerable impact worldwide. With the outbreak starting from Wuhan, China, within the span of less than a year, it has spread across the globe and effectively shutdown most countries. However, it could be observed that numerous researchers have conducted various studies related to this topic in a short period of time. Particularly, focusing the country the virus outbreak began, China. 
The analysis of social media related to this outbreak was one such way the behaviour of people during this crisis was analysed. The study by Zhu et al. \cite{zhu2020:CORONAChinaWeebo} presents one such example where randomly sampled Weibo users from China, were analysed to identify the level of the attention the users had towards the COVID-19 virus. 
The work by Gao et al.\cite{Gao2020} discusses about the mental health issues which were highlighted in social media during COVID-19 pandemic in China.
In addition, they have described how exposure to social media caused the development of mental health problems during the crisis period.
Furthermore, the research by Negin Karisani and Payam Karisani  \cite{karisani2020mining} analysed the impact of the outbreak in China based on data from Twitter. In this research, authors have used state-of-the-art machine learning models to determine the disease spread based on insights gained from social media.

In addition, representing the spread of social media messages similar to a disease, Cinelli et al. \cite{cinelli2020covid19:italypaper} analysed messages related to the COVID-19 outbreak and presented patterns as to which they spread in a variety of social media platforms. Furthermore, the research by Boberg et al. \cite{boberg2020pandemic:germanpaper} analysed Facebook pages of alternative news media and presents yet another example for studies conducted on social media that relates to the COVID-19 outbreak.
The research by Murray et al. \cite{murray2020symptom} analysed the COVID-19 outbreak from the perspective of personal experiences utilising data from the social media platform, Reddit. In this study, the authors used social media data to generate important insights on symptoms of the disease as well as sentiments felt during the outbreak.
Kadam and Atre \cite{Kadam2020} discussed the panic that ensured in social media during COVID-19 outbreak in India. The authors analysed social media content and their corresponding societal responses to identify their negative impact to the measures taken to curtail the virus outbreak in the country. 
The study by Depoux et al. \cite{Depoux} stands out in the aspect on how it portrays social media as a pandemic that spreads misinformation and rumours greater than the speed of transmission of COVID-19.
Ziems et al. \cite{ziems2020racism} presents a research on how racism was developed on social media during the COVID-19 outbreak. Here, they have specifically focused on how racial hate towards Chinese and other Asian countries spread, and role of counter-hate speech on mitigating the spread.
Major social and behavioural changes of people because of the massive global health crisis were analysed by Bavel et al. \cite{Bavel2020}.
In addition, the authors have evaluated how psychological and behavioural science concepts could be used to mitigate the negative impact of the health crisis.

\section{Methodology}

\subsection{Data collection and Categorisation}

The research conducted focused on using Facebook as its primary source of data. 
This specific selection of Facebook was due to it being the most popular medium of social media used in Sri Lanka. The extent of the popularity was such that the user-base of Facebook was multiples of times larger than other social media platforms combined  \cite{Statcounter:SL}. Thus, it was the belief that data collected from Facebook would lead to a sample that would better represent the Sri Lankan populace.  

Following the decision regarding the social media platform to collect the data, data of posts from 50 sources were manually collected and categorised. These sources were comprised of both profiles as well as pages that were selected at random. 
However, the persons in charge of such sources spanned across a wide array of ideals as well as professions. In addition, it was assumed that for profiles, an individual was in charge, and for pages, a group of like-minded people were in change.
A notable limitation regarding this method would be that it would only capture the behaviour of the individuals or groups that are in charge of these 50 sources and not the entire populace as a whole. 
The reason for employing such small limits and manual collection mechanisms was due to the restrictions placed in the Facebook platform to extract data. 
However, as the collective number of posts amounted to 2398, we believe that the sample size would be sufficient for this exploratory study. Nonetheless, there were certain added benefits of this method in comparison to a standard data collection method via a query. Such benefits were that this method would also take image posts such as memes as well as posts that do not directly refer to the COVID-19 virus but were posted due to an event related to it into account as well.

\begin{figure}[t!]
    \includegraphics[width=\columnwidth]{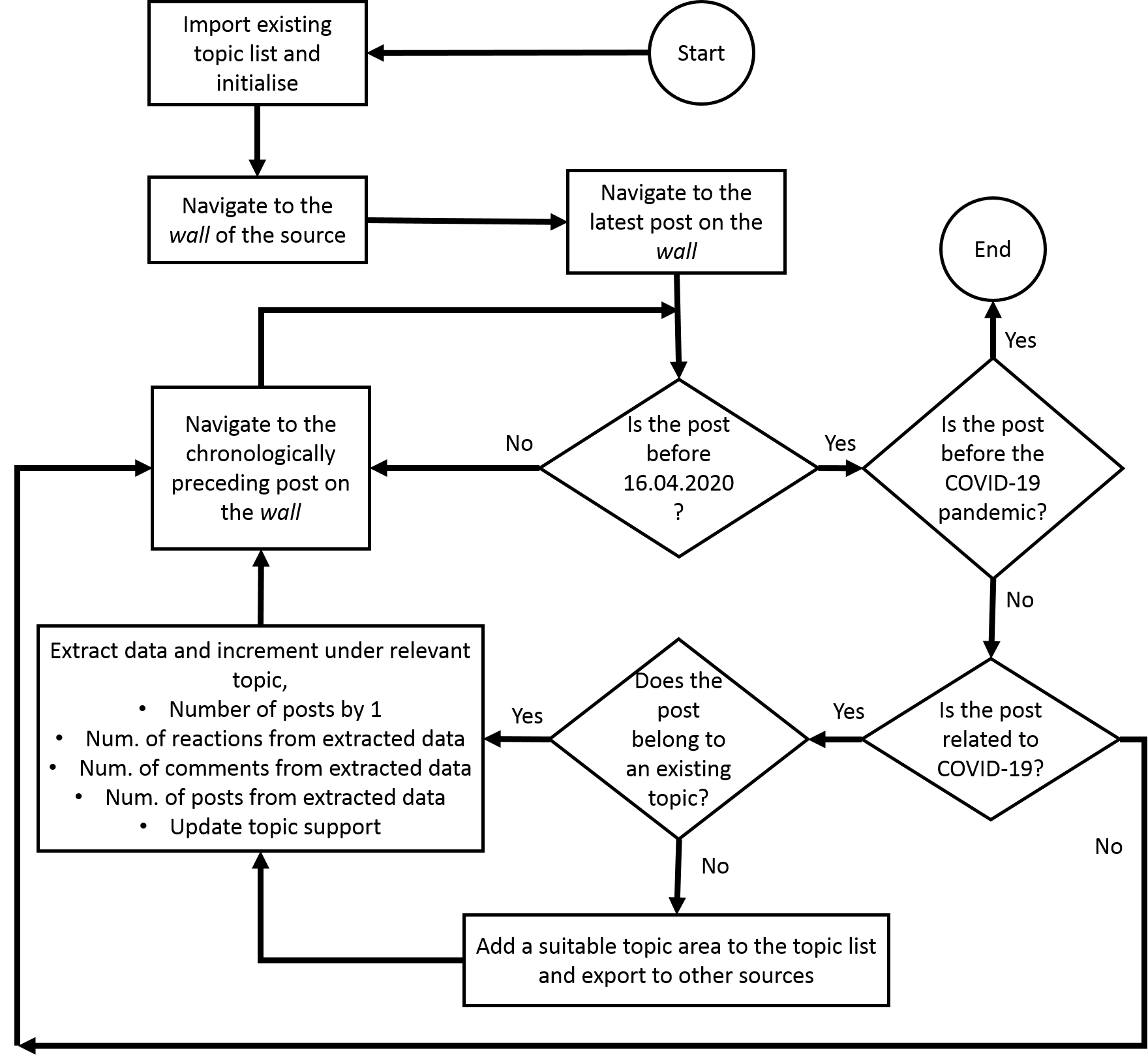}
    \caption{Flowchart of the data collection process of a single source}
    \label{flowchart}
\end{figure}

In the data collection process itself, public posts along with data regarding its significance were manually collected that related to the COVID-19 virus or originated due to the result of the outbreak. To elaborate, taking a source at a time, the \textit{wall} of the source was scanned for related posts. Once found, each post was then counted for one out of 57 categories along with its the number of the reactions, number of comments as well as the number of shares. It is these numerical data that will be used to measure the significance of posts. In addition, an attribute was stored to track if a source posted the content that either supported or criticised a particular topic. However, if multiple posts contradicted one another, it was taken as that source neither supported nor criticised the topic in question. A profile that first criticised the actions of the government and then later praised them would be an example of such a situation. Furthermore, it should be noted that the initial 57 categories were created empirically during the categorisation process itself if a post found did not fit into any of the existing categories. To give an example, a post detailing the negative effects of panic buying was found in one of the sources. The list of topics would then be checked, and if a suitable topic such as \textit{Panic buying} was present, details of the post would be added under that topic section for that particular source. If such a topic did not exist till that point, then it was created and added to the list of topics. Following such action, the details of the post would then be marked under this new topic section. Fig. \ref{flowchart} depicts the aforementioned process for a single source.

Following the completion of the data collection, the initial categories were reduced to 38 by combining related topic areas. The reason for this action was to eliminate the redundancy that was introduced when collecting data as certain topics areas were more specific instances of another more general topic area. The final set of topics utilised in the categorisation process as well as the cluster they belong to are mentioned presented in Table \ref{table:topics}.

\begin{table*}[]
    \caption{Topic clusters and topic categories}
    \centering
      \begin{tabulary}{\textwidth}{C||LL}
        \hline
        \hline
        \textbf{Topic Cluster} & \multicolumn{2}{c}{\textbf{Topics}}
        \\
        \hline
        \hline
        \multirow{4}*{Start of the outbreak} & \tabitem Situation in China &  \tabitem Other things to worry about / Other diseases
        \\
        &  \tabitem Early cases in Sri Lanka &  \tabitem Bringing people back / Support for those in other countries
        \\
        &  \tabitem Curing 1st case in Sri Lanka  &  \tabitem Early measures to stop the virus
        \\
        &  \tabitem Irresponsible behaviour &  \tabitem People coming back by themselves from other countries 
        \\
        \hline
        Quarantine by Govt & \tabitem Quarantine by the Government
        \\
        \hline
        Lock-down & \tabitem \#LockdownSL & \tabitem Government decisions for social distancing
        \\
        \hline
        \multirow{3}*{Supplies} & \tabitem Panic buying & \tabitem Discipline after curfew release
        \\
        & \tabitem Shortage of supplies / services & \tabitem Deficient / Overpriced products
        \\
        & \tabitem Delivery & \tabitem Local production / inventions
        \\
        \hline
        Economy & \tabitem Economy & \tabitem Government's economic decisions
        \\
        \hline
        \multirow{2}*{Politics} & \tabitem President / Government & \tabitem Other political parties / Politicians
        \\
        & \tabitem Postponing election & \tabitem Media / Social Media
        \\
        \hline
        \multirow{2}*{Advice} & \tabitem Scientific / Valid advice & \tabitem Cultural and Religious beliefs / activities
        \\
        & \tabitem Rumours & \tabitem Information about Sri Lanka
        \\
        \hline
        \multirow{3}*{Services / People} & \tabitem Armed forces & \tabitem Religious / Famous figures / Top companies 
        \\
        & \tabitem International organisations & \tabitem Other essential services / All services
        \\ 
        & \multicolumn{2}{l}{\tabitem Medical / Government Medical Officers' Association (GMOA)}
        \\
        \hline
        Humour & \multicolumn{2}{l}{\tabitem Humour / Entertainment / Online activities / Boredom relief}
        \\
        \hline
        Donations & \tabitem Donations / Helping others / Needy
        \\
        \hline
        Global Situation & \tabitem Situation in other countries / World
        \\
        \hline
        \multirow{2}*{Other} & \tabitem Benefits of the COVID-19 outbreak & \tabitem Explain via Maths  / Curves
        \\
        & \tabitem Dire situation / Sorrow
        \\
        \hline
        \hline
      \end{tabulary}  
    \label{table:topics}
\end{table*}

Furthermore, the data collection process only considered Facebook posts content created on or before the 15\textsuperscript{th} of April 2020. This date, in particular, was kept to ensure that the data extracted from each source spanned through a consistent period of time. As this research primarily focuses on the initial reactions to the COVID-19 outbreak, we deemed that the aforementioned date would be sufficient as it was roughly a month after the virus outbreak in Sri Lanka.  

The aforementioned collection process was carried out by a team of individuals who were significantly knowledgeable in data analysis. 
Although the data collection process was cross-validated among the team members, validation of the data itself was not attempted due to the challenges from the dynamic nature of posts in Facebook. 
However, to mitigate the impact of fluctuations in the data, the collection process was initiated approximately a week after the cutoff date for considering posts. In addition, due to any privacy issues that could arise, the collected data was kept secured and was not included in a public domain.

\subsection{Normalisation}
An observed point of imbalance in the data is the vastly varying number of followers of each source. Due to this variance, a significant difference could be observed considering the number of reactions, comments and shares from once source to another. When considering the total significance per topic, this factor would skew the data in the same pattern or behaviour of the more popular sources than considering all the sources generally as a whole. Thus, the attributes related to significance, such as the topic-wise number of posts, reactions, comments as well as shares were subjected to normalisation considering each person.   

\begin{equation}
\hat{s}_{k}^{i} = \frac{100 \times s_{k}^{i}}{\sum_{n=1}^{38}s_{n}^{i}} \label{eq1}
\end{equation}

The task of normalisation was accomplished by dividing a significance attribute of a topic by the sum of the said attribute's value for all the topics of the person considered. The result was then multiplied by 100 to get a more significant figure to avoid small fractions. The equation relating to this process is given by \eqref{eq1}. 
Here, \textit{s$_{k}^{i}$} denotes some significance attribute such as the number of posts, reactions, comments or shares pertaining to some topic for the i\textsuperscript{th} source, while \textit{\^s$_{k}^{i}$} denotes its normalised value. 
Furthermore, \textit{k} denotes the topic that is taken into consideration while \textit{n} iterates over all the topics. To give an example, assume that a profile has gathered 400 likes for posts in the \textit{Panic buying} topic category. If the said profile had gathered 2000 likes in total, the normalised number of likes for the \textit{Panic buying} topic would be 20 for that profile.

\section{Observed patterns of behaviour}
Based on the collected data as well as key events observed throughout the relevant time period, various patterns were identified in regards to the behaviour of the Sri Lankan populace. In addition to the data collected specifically for this research, certain observations were also taken into account from screenings which were randomly timed to support the patterns identified. 

Furthermore, this analysis of behaviours was conducted focusing on generalised key topics that were prevailing at the time. This was done in order to conduct a more in-depth analysis, where each key topic contains subtopics that are related to one another. Although there is an overlap between such subtopics, the underlying intention behind them is in relation to the main topic. For example, posts related to key topic areas such as returnees from foreign countries as well as resource acquisition may both include posts that fall under the \textit{President / Government} category.

The behaviours patterns observed for each sub-topic are illustrated in Fig. \ref{flow:1}, Fig. \ref{flow:2} and Fig. \ref{flow:3}. Here, the solid arrows represent the majority pattern observed in public, while the dashed arrow represents the minority. It should be noted that the public observations were taken from the random scans, and do not necessarily follow the same patterns as the collected data. The topics, along with the reactions observed to the topics, are represented by the rectangular boxes. These reactions were either support or against. The darkly shaded boxes with solid outlines represent that the reaction to that topic is the majority reaction in the data. The opposite of this is represented by the lightly shaded boxes with dashed outlines. The white boxes represent that the reaction for the topic that it contains is neutral in the data. In addition, the boxes that connect to an arrow present a plausible reason for the flow pattern represented by that arrow.

\subsection{\#LockdownSL}
	
\begin{figure}[h!]
    \includegraphics[width=\columnwidth]{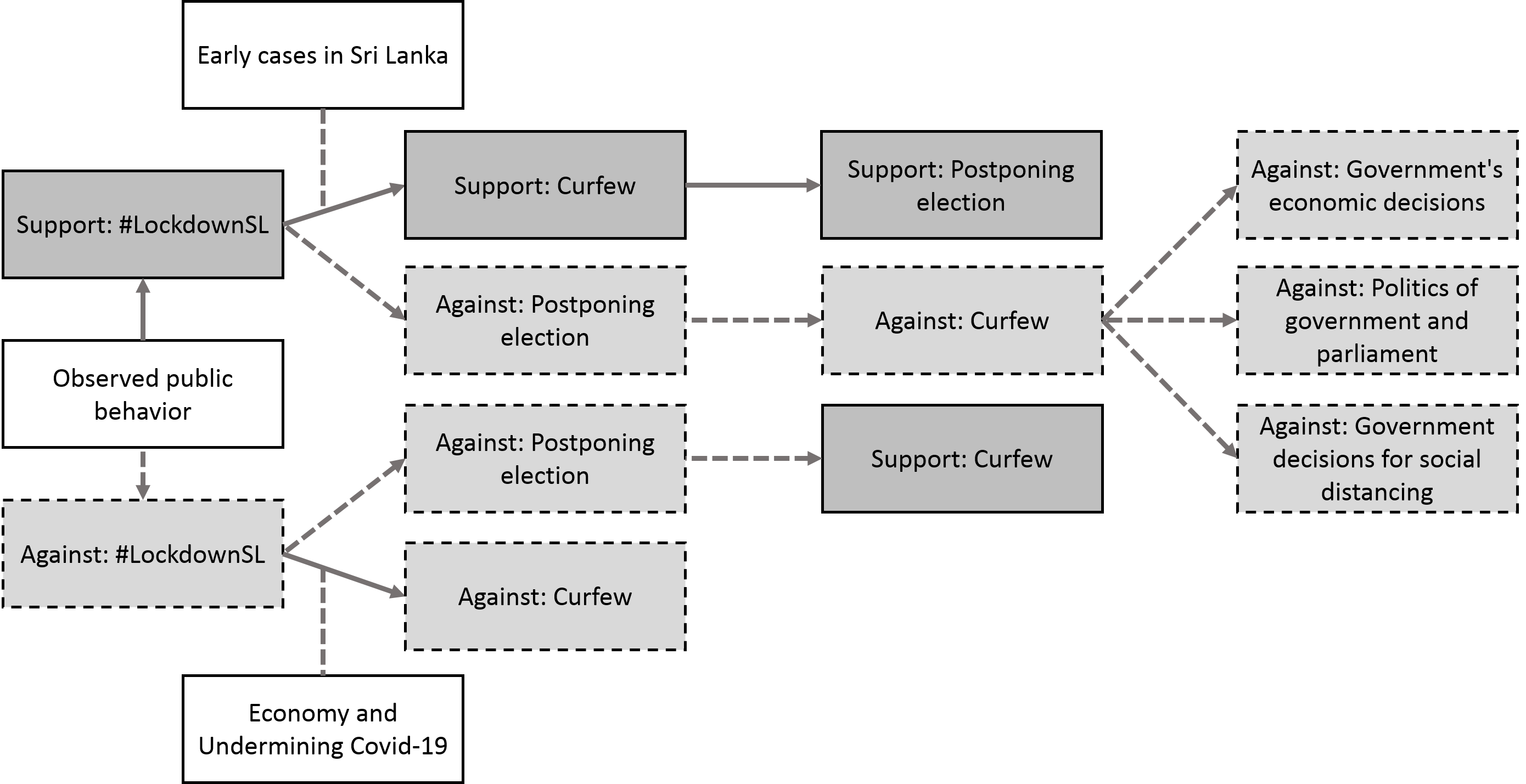}
    \caption{Flow diagram showing patterns related to the lock-down}
    \label{flow:1}
\end{figure}

This can be considered as one of the first major movements that provoked the majority of the Facebook audience in Sri Lanka to post content related to the COVID-19 pandemic. The pattern regarding the flow of post categories related to this topic is presented in Fig. \ref{flow:1}.

Well after the recovery of the first COVID-19 patient who was a foreign nationalist, the identification of the second case in Sri Lanka sparked panic among the people. At this point, the government was constantly requested to lock-down the country to prevent outsiders who have contracted the virus from coming in. The public did so on Facebook by changing their profile pictures and posting with the hashtag \textit{\#LockdownSL}, demanding the government to go into lock-down.
As the second patient contracted the disease due to being in a profession that regularly associated with foreigners, the call to lock-down the country could be somewhat justified. In addition, due to the fact that this was the first case of COVID-19 being contracted to a Sri Lankan nationalist within Sri Lanka, may have led the public to overreact.  

Although the majority of the population supported the Lock-down campaign, some stated that it was not required with the at that time circumstances. An argument presented by such people was that it was too early to take such strict action. In addition, due to Sri Lanka being a country where international tourism plays a great role in its' economy, another argument that could be seen against the lock-down was that such action would not be bearable for the country.

With such observations, a deeper analysis was done on the behaviour the people who both supported and rejected the lock-down campaign via their Facebook profiles.
From the people who stood with imposing the lock-down, the majority seemed to be discussing the initial cases found in the country.

Additionally, some people were talking about the lock-down to indirectly support their political preferences. This was especially highlighted during the heated discussion that occurred on the decision to postpone the general election. 
To give some insight into this matter, due to the Presidential election in 2019, the government that existed before it lost power as the prevailing president was primarily supported by the opposing party.
Thus, the prevailing government is that of the said opposing party. In order to create a parliament that supports the prevailing president, the holding of the general election is pivotal. Thus, the goal of the prevailing government was to hold it at the earliest. 
Out of the people who discussed about the lock-down, the majority made their next few posts supporting or apposing various political decisions. This polarised behaviour was expected as these series of incidents had a clear influence on the populace based on their political preferences.

Additionally, the government also took immediate action after the initial COVID-19 cases were discovered. These actions taken had the foreboding signs that a curfew would be placed and the country would be truly locked-down. Thus, it came to no surprise when an island-wide curfew was later announced by the government. 
However, this curfew lasted for an unprecedented amount of time. Low-risk areas were given intermittent relaxing of restriction, while the capital and nearby areas experienced a complete lock-down surpassing 70 days due to the high risk of community spreading. 
To make things intense this kind of civil liberty restrictions were not experienced even during the intense times of the 30-year war witnessed by the Sri Lankans.
When considering the posts after imposing the curfew, the majority of the people who participated in the \textit{\#LockdownSL} movement, again supported curfew. This again was no surprise as this was the practical implementation of their wish. However, there were also some who supported the lock-down but were against the curfew. This may have been due to the intense restrictions that came about due to the curfew, which were not apparent initially. Nonetheless, some of these individuals also appeared to have found fault no matter what action the government took. However, the people who stood against the lock-down appeared to have supported the curfew.

Thus, considering the patterns and behaviours, it is possible to hypothesise that certain individuals who posted content relating to the lock-down of the country did so following a trend. Such people had no proper idea about the major concerns or requirement of the lock-down. However, it could also be possible that some of these individuals had second opinions regarding their initial decisions.

\subsection{The return of those from foreign countries}
\begin{figure}[h!]
    \includegraphics[width=\columnwidth]{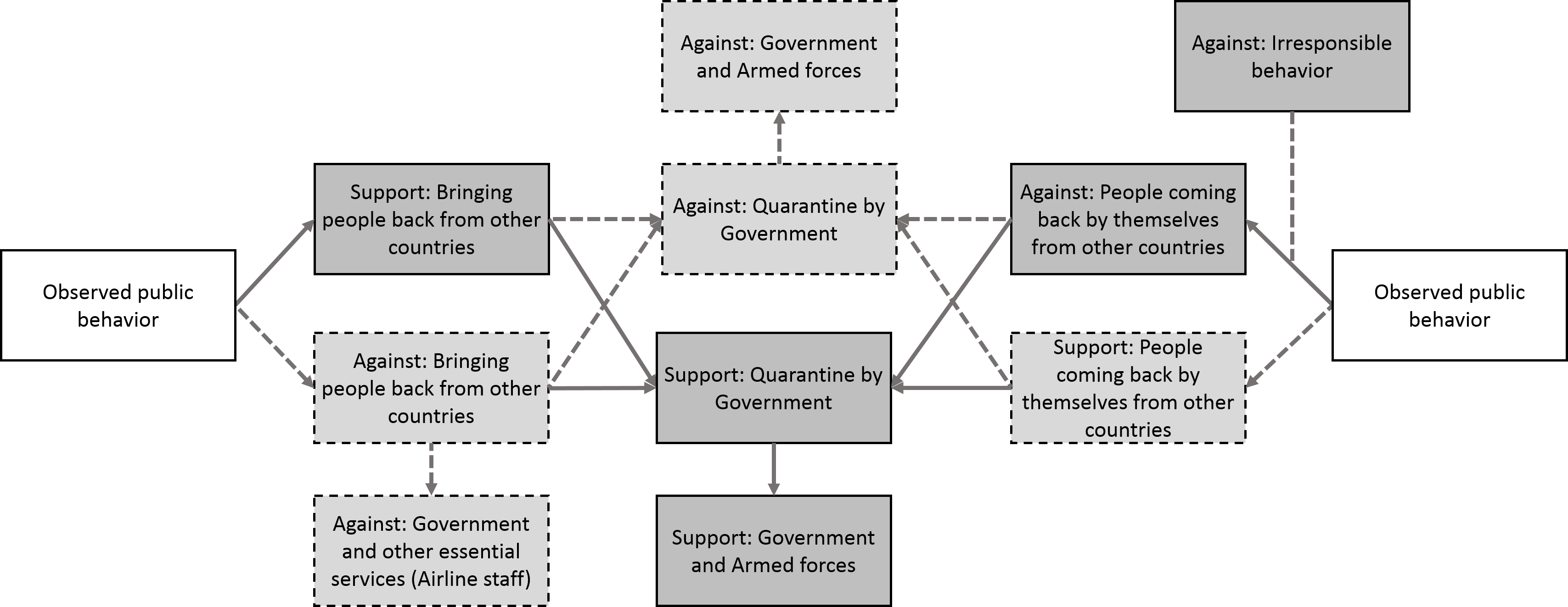}
    \caption{Flow diagram showing patterns of foreign returnees of Sri Lanka}
    \label{flow:2}
\end{figure}
This topic focused on patterns from Facebook posts of the Sri Lankan populace that related to the return of those from foreign countries. Observed flows of sub-topics related to this topic are presented in Fig. \ref{flow:2}.
Content related to this topic can be seen across two main phases.
The first phase began with the news of the students who were stranded in Wuhan, the birthplace and epicentre of the COVID-19 outbreak.
With this discovery, many people, especially the relatives of the students requested the government to bring the students back. In Facebook, it was observed that the majority of the public supported this request and the cause of bringing the students back. 
However, a minority stood against it by highlighting the fact that there is a possibility the students could carry the virus and spread it in Sri Lanka. 
Nonetheless, the government sent an aircraft to fetch the students. 
After the returning to Sri Lanka, the students were directed to a quarantine facility which was coordinated and run by the armed forces. It could be observed that the majority supported and praised the government as well as those who were involved in this action. However, there was a minority that stood against this decision. Furthermore, there were minor reports that the airline staff and their family members were discriminated.

The second phase initiated with the arrival of returnees from countries that were severely hit by the COVID-19 virus.
During this time, it was observed that the general public was more aware of the severity of the virus and they were more hesitant to support the people coming back from foreign countries. This was regardless of the fact that the government had decided to maintain the quarantine process for such arrivals.
In addition, during this time period, an incident occurred where returnees from a country where the COVID-19 virus was rampant protested against the quarantine process at the airport. The neglection of health advice by these individuals was a key point of that incident.
With this, the aforementioned public hesitance turned into outright resistance. Thus, the majority of the Sri Lankan populace turned against such returnees. However, as time progressed, all arriving from foreign countries were subjected to the quarantine process regardless of the county they arrived from.

It was observed that this quarantine process and the armed forces who carried it out were heavily praised and supported by the majority of the public. Although there was also a minority that undervalued and criticised such efforts.

Thus, considering these observations, it is possible to conclude that the public was initially supporting those stranded in other countries. However, due to the detrimental actions of those who arrived later spreading on Facebook, this support turned into opposition. Nonetheless, the majority of the public continued to support those were in essential services throughout the entire period. 

\subsection{Resources and their acquisition}
\begin{figure}[h!]
\includegraphics[width=\columnwidth]{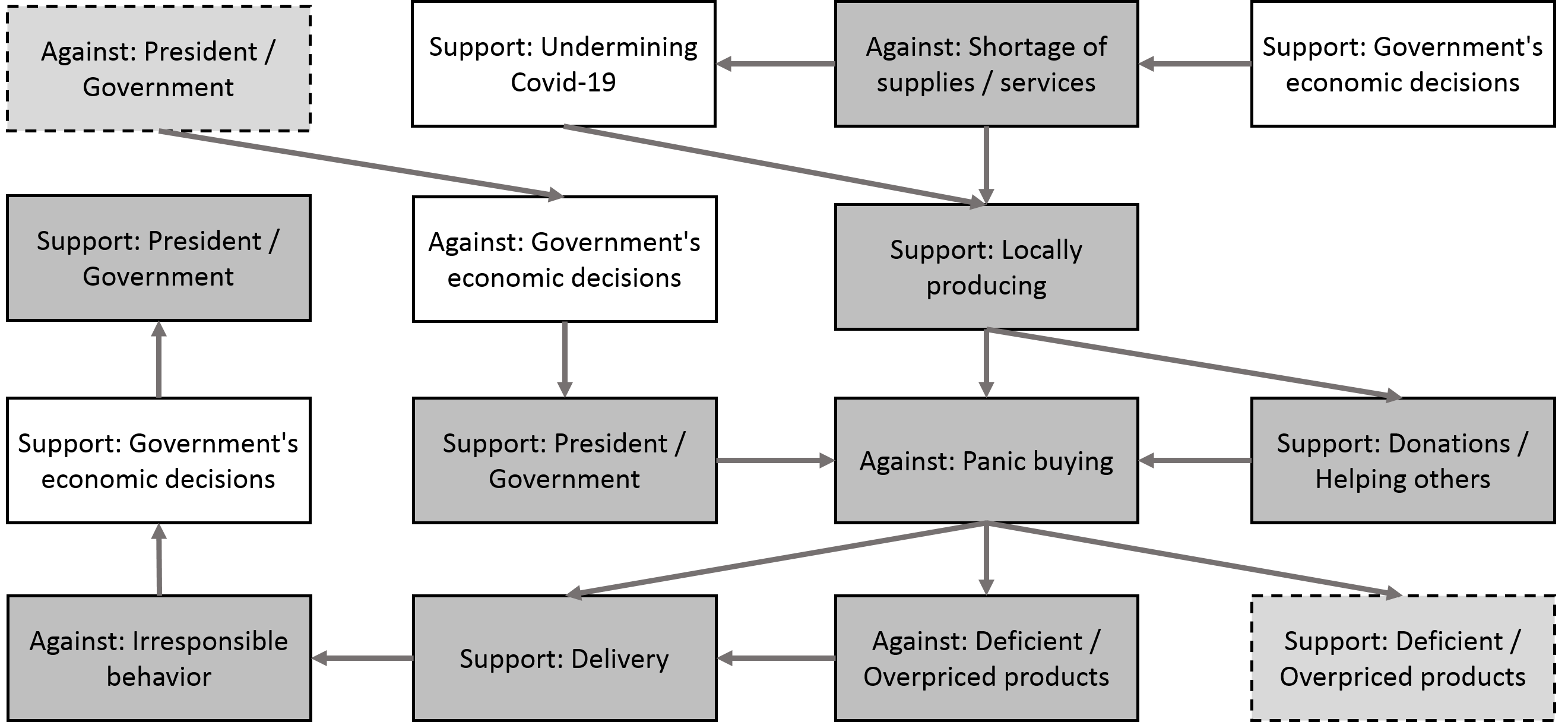}
\caption{Flow diagram showing patterns of resources and their acquisition}
\label{flow:3}
\end{figure}
The topic flow patterns which were related to this topic are presented by Fig. \ref{flow:3}. Initial posts that fell under this section were focused on the various decisions by the government of Sri Lanka that related to its economy.
One such decision, in particular, was not lowering the oil prices in the country even though the price of crude oil was reduced significantly.
Though it is to be expected for people to lash out against this matter, certain individuals defended this decision. 
These individuals later supported their standpoint when it was revealed by the government that the extra monetary resources collected by not lowering oil prices, were used to lower the price of certain food items. 
Specifically, these food items were canned salmon and dhal. It was clear that this decision was taken to reduce the burden of low-income families in the country if the country went into lock-down. 
However, the presence of political influence regarding posts that either supported or criticised such government decisions could not be denied based on the older posts of the sources.

Starting from initial stages of the outbreak, posts admonishing the act of panic buying were regularly observed. This was especially true after the curfew was announced. 
However, such posts did not portray the severity as in certain other countries where shelves were barren and devoid of goods.
Instead, most of these posts focused on the act of crowding in large queues when shopping. The argument was that this was counter-intuitive as it would help spread the virus more rapidly.

Following the island-wide curfew by the government, posts relating to helping the needy could be commonly seen. Such posts pointed out that groups such as the destitute, elderly and orphans would be severely affected by their lack of means to get food for their daily lives. As if a response to these, multiple posts showing food donations to such vulnerable groups and even stray animals could also be observed.

Following the lifting of the curfew in districts without a major spread of the virus, the appearance of posts praising the discipline of Sri Lankans for properly following the necessary protocol began to emerge. This was due to such people properly maintaining a distance of 1 metre when in queues.
However, when the curfew was lifted in more densely populated districts the following day, crowding at stores could be witnessed. This caused posts criticising panic buying and crowding to reemerge. Nonetheless, such crowding may have been inevitable as the people in such dense areas would have been starved for essential resources with only a limited number of outlets to acquire them.   
In addition, posts could be found showing those whose professions were essential to combat the virus such as nurses getting into long queues at stores. These posts also requested that such individuals be given preference when in queues as they were the vanguard in combating the virus.

However, as the government came to realise that the avoidance of such crowding was implausible, the country was put into indefinite lock-down. Following this course of action, the promotion of delivery mechanisms to distribute essential goods was looked into. 
Due to this, posts relating to such deliveries began to circulate in social media. A type of post, in particular, that could be observed was where people would request for contact numbers of essential goods deliveries as well as services. This indicated the lack of a proper mechanism to obtain methods of contacting crucial services. 
However, due to the apparent unethical practices by certain companies when making deliveries, posts denouncing such companies emerged.
In addition, a surge of posts promoting local productions as well as innovations also sprang forth sporadically. Such posts heavily promoted nationalism stating that Sri Lanka should stop relying on imports and focus on producing required goods on its own.

Thus, the majority of the posts found relating to this topic area seemed to focus on advising on how to properly acquire resources. The method of doing so took on many forms such as pointing out facts which were oblivious or otherwise and even direct criticism.

\section{Result analysis}
Following the categorisation of the posts from the sources considered, various forms of analyses were conducted in order to identify key points relating to the behaviour of the selected set of sources.

\subsection{Topic-significance}

Based on the data collected from 50 profiles and pages, an analysis was performed to identify the topics that were highly discussed on Facebook throughout the period of study. 
As Facebook appeared to have more than 70 percent of the total social media user-base in Sri Lanka\cite{Statcounter:SL}, it could represent the majority of the general public better than any other social media platform. Thus, this analysis portrays the significance of the various discussions that occurred in society during the pandemic situation in Sri Lanka.

As described in the methodology section, following the generation of the 38 topic categories the results were subjected to normalisation based on each profile. From these normalised values, \eqref{eqtts} was derived to calculate the total significance per topic.

\begin{equation}
T_{k} = \sum_{i=1}^{50}[\hat{s}_{k,p}^{i}+\hat{s}_{k,r}^{i}+\hat{s}_{k,c}^{i}+\hat{s}_{k,s}^{i}]
\label{eqtts}
\end{equation}

In  \eqref{eqtts}, \textit{T\textsubscript{k}} denotes the total topic significance for \textit{k\textsuperscript{th}} topic. In addition, \textit{\^s$_{k,p}^{i}$}, \textit{\^s$_{k,r}^{i}$}, \textit{\^s$_{k,c}^{i}$}, as well as \textit{\^s$_{k,s}^{i}$} corresponds to the \textit{i\textsuperscript{th}} person's normalised post, reaction, comment, and share counts.

\begin{figure*}
\includegraphics[trim={0 1cm 0 3cm}, width=\textwidth]{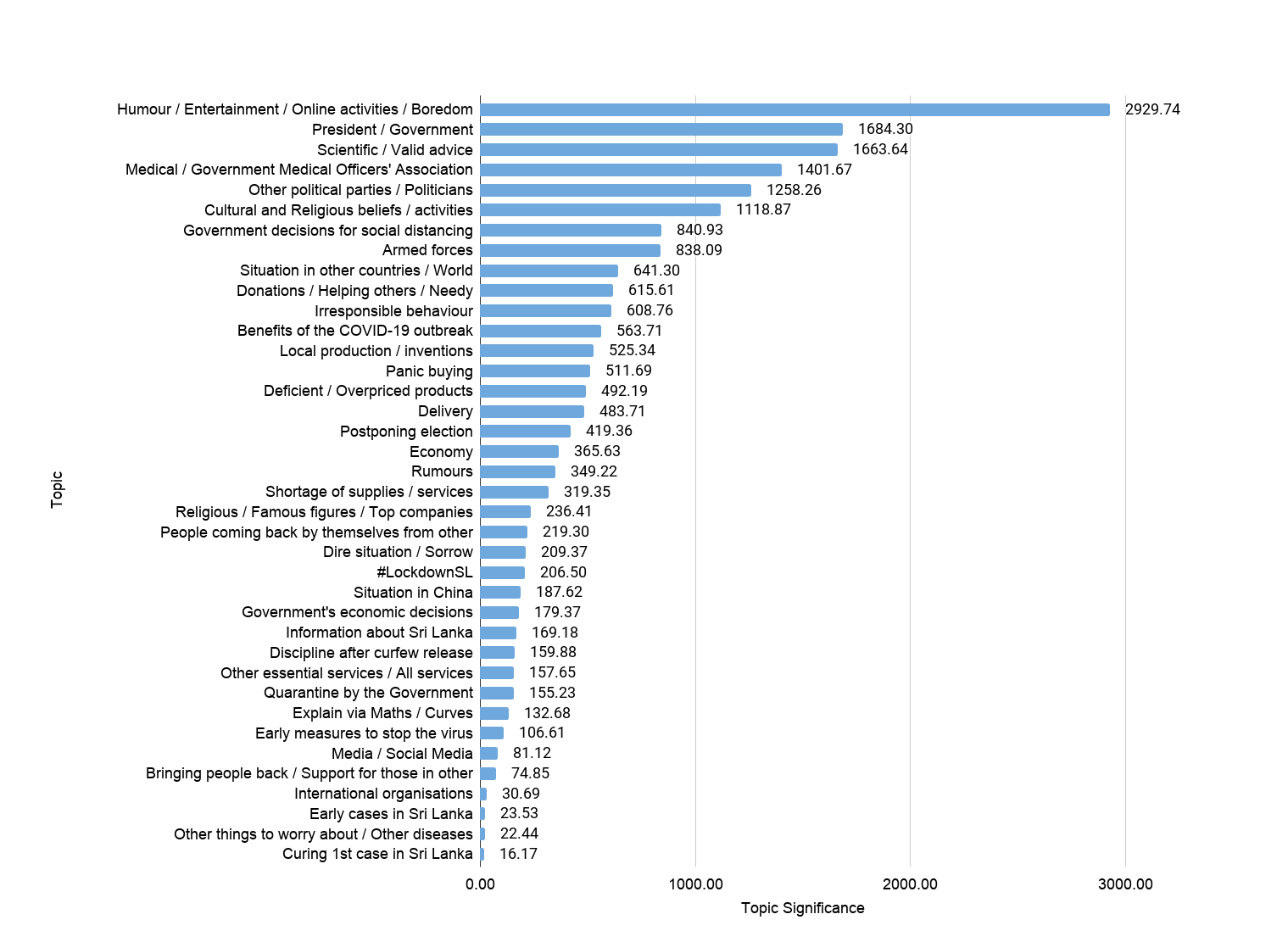}
\caption{Total significance of each topic}
\label{fig:bar}
\end{figure*}

The bar graph in Fig. \ref{fig:bar} illustrates the significance of the major topics that were discussed during the period of the research. Here, the 38 different topics were ordered based on the topic significance. A significance score ranging from 2929.74 to 16.17 could also be observed. 

According to the bar graph in Fig. \ref{fig:bar}, the topic cluster of \textit{Humour / Entertainment / Online activities / Boredom relief} had the highest topic significance score of 2929.74, surpassing all other topics by a considerable margin.
This behaviour could be justified as there were plenty of sarcasm based posts encountered throughout the study.
It seemed that the majority of the people appeared to have used various incidents that happened during the period to be making posts on humour and entertainment. With the policies to ensure social distancing in place, the sensation of solitude that arose because of it may have led the people to use Facebook as a boredom releasing mechanism.

Apart from humour, another prominent area was politics, where several topics with high significance scores could be observed.
This situation was also expected as there was a parliamentary election which originally happened to be held on 25\textsuperscript{th} of April. Thus, most of the populace were discussing political related topic groups such as \textit{President / Government} (1684.30), \textit{Other political parties / Politicians} (1258.26), \textit{Government Decisions for social distancing} (840.93), and \textit{Postponing election}(419.36).
However, those who discussed such topics interpreted political matter in a very polarised manner that either supported or opposed the government. 

The topic \textit{Scientific / Valid advice} which held a score of 1663.64, indicated that Facebook was also being used as a medium to exchange valid and scientific advice to mitigate the spread of the virus. 
The influence of famous personalities may have been the cause of this behaviour. This was due to such influential figures posting videos on how to properly behave during the COVID-19 outbreak, and included individuals such as cricket players, entrepreneurs, journalists as well as politicians. This, in particular, can be highlighted as a promising use of social media during a crisis.

During this disastrous period, there were several entities and authorities that played a splendid role in controlling the virus from spreading in the country. 
Thus, Members of the Government Medical Officers' Association (GMOA), other medical officers, as well as armed forces were held in high regard by the people. This made posts related to such topics become especially highlighted.

When moving down the graph, a cluster of topics with medium impact where the significance varies from 641.30 to 319.35 could be seen. These topics mainly covered the acquisition of essential resources as well as the various behaviours of the people.

Topics such as \textit{Early cases in Sri Lanka}, \textit{Situation in China}, \textit{Early measures to stop the virus} and \textit{Explain via Maths / Curves} had a lower significance when considering topic scores. Most of these topics were discussed in the initial period of the pandemic, especially before the government took strict actions.
Thus, it indicates that people appeared to be less interested in the virus during the early stages.
Generally speaking, when comparing the topics against time, it was observed that most of the topics came into the discussion after the lock-down of the country. Additionally, significant incidents such as panic buying as well as returnees from foreign countries, bolstered the attention people had towards COVID-19.

\subsection{Heat-map}

In order to get a holistic idea on the relations between topics, profiles and significance, a heat map was derived. In addition, the topics were clustered further to give a more compact representation.

\begin{figure*}
\includegraphics[width=\textwidth,height=0.2\textheight]{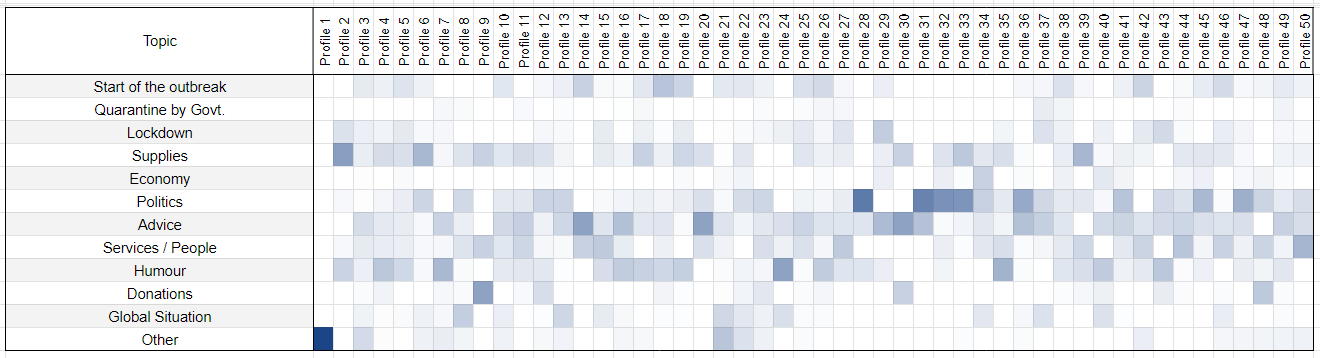}
\caption{Heat-map showing profile-topic-significance}
\label{fig:heat}
\end{figure*}

The heat map in Fig. \ref{fig:heat} presents 12 topic clusters that were made from the 38 initial topics. The columns represent the different sources used to make the data sample. Each cell had been coloured based on its significance, where the higher the significance, the darker the colour. The significance has been calculated per profile utilising  \eqref{eqppts}, which was based on the normalised number of posts, reactions, comments and shares.

\begin{equation}
T_{k}^{i} = \hat{s}_{k,p}^{i}+\hat{s}_{k,r}^{i}+\hat{s}_{k,c}^{i}+\hat{s}_{k,s}^{i}
\label{eqppts}
\end{equation}

Here, \textit{T$_{k}^{i}$} denotes the \textit{k\textsuperscript{th}} topic's significance of the \textit{i\textsuperscript{th}} source. Similarly,
\textit{\^s$_{k,p}^{i}$}, \textit{\^s$_{k,r}^{i}$}, \textit{\^s$_{k,c}^{i}$}, and \textit{\^s$_{k,s}^{i}$} corresponds to the \textit{i\textsuperscript{th}} person's normalised number of posts, reactions, comments, and shares.

\subsection{Topic Co-Occurrence}
In addition to the heat-map, as a means of analysing the interrelation of the topics, their co-occurrences were also calculated.
This was accomplished by first taking any two topics, and then counting of the number of profiles or pages from the sources that contained both of the topics. The equation used when considering two sources was \eqref{eq:cooc}, where the value of \textit{c$_{k,l}^{i,j}$} denotes if the topics \textit{k} and \textit{l} co-occurred between the \textit{i} and \textit{j} sources. 

\begin{equation}
  c_{k,l}^{i,j}=\begin{cases}
    1 & \text{i and j posted content on topics k and l}.\\
    0 & \text{i or j did not post content on k and l}.
  \end{cases}
  \label{eq:cooc}
\end{equation}

These calculations between each unique pair of sources were then summed to get the final co-occurrence scores between topics as in \eqref{eq:tcooc}. Here, \textit{C$_{k,l}$} denotes the total co-occurrence score between the topics \textit{k} and \textit{l}. Although unsurprising, the pair of topics with the highest co-occurrence count of 31 was the \textit{Humour / Entertainment / Online activities / Boredom relief} and \textit{President / Government} topic groups, which also boasted the highest normalised significance values.

\begin{equation}
  C_{k,l}= \sum_{i=1}^{50}\sum_{j=i+1}^{50}c_{k,l}^{i,j}
  \label{eq:tcooc}
\end{equation}

As a means of analysing the co-occurrence of topics, the co-occurrence graph in Fig. \ref{fig:fullcoo} was created.
In this graph, the 38 topic categories were represented as nodes. 
Furthermore, to associate co-occurrence with the topic frequency, a score was created using the normalised sum of posts and then represented in the graph via the size of the nodes. 
The nodes were then ordered in a circle using this attribute which was calculated by using \eqref{eq:freq}. Here, \textit{f$_{k}$} denotes the score for topic \textit{k}, with the remaining symbols defined by the notations mentioned previously.

\begin{equation}
  f_{k}= \sum_{i=1}^{50}\hat{s}_{k,p}^{i}
  \label{eq:freq}
\end{equation}

In Fig. \ref{fig:fullcoo} the node with the highest popularity is located at the top with the sizes of nodes decreasing counter-clockwise.
In addition, the edge weights of the graph, which are also represented by the thickness of the edges, utilise the co-occurrence values calculated previously.

\begin{figure*}
\includegraphics[trim={0 3cm 0 3cm}, width=0.8\textwidth]{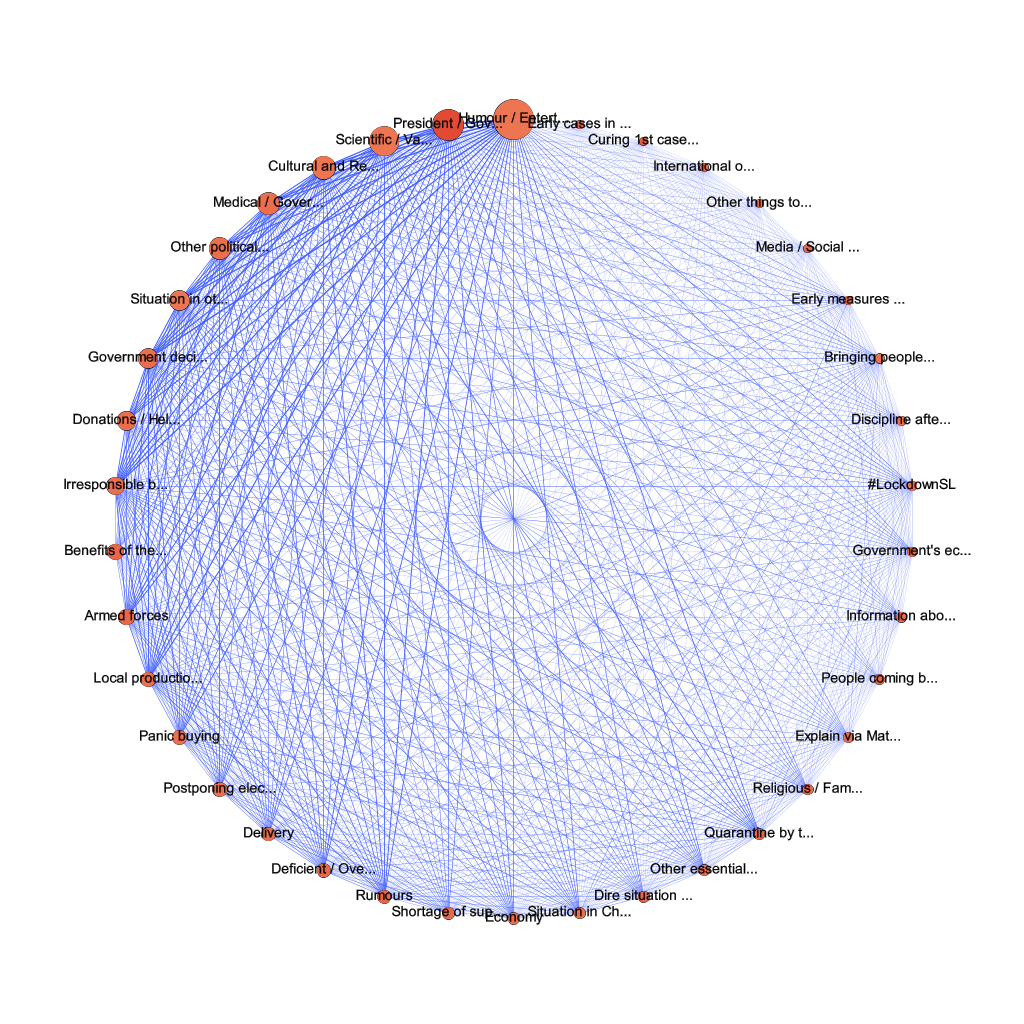}
\centering
\caption{Co-occurrence graph of the topic categories based on data sources}
\label{fig:fullcoo}
\end{figure*}

Upon analysis of Fig. \ref{fig:fullcoo}, it can be observed almost every topic has a co-occurrence with one-another.
This is presented by the existence of interconnecting edges between almost each and every node.
There were only 6 pairs where a co-occurrence could not be observed. 
Out of these 6, 5 had one of the two least significant topics as one of the topics in the pair. These topics were \textit{Other things to worry about / Other diseases} and \textit{Curing 1st case in Sri Lanka}.
Furthermore, in general, the less frequent topics had a reduced level of co-occurrence compared to their more popular counterparts, in terms of the number of sources the co-occurrence occurred.
The rationale behind this deduction is that due to the upper right section of the graph showing a fairly lighter colour while the upper left section showing a fairly darker colour. This was due to the cumulative colours of the weighted edges.
Through such observations, another possible deduction would be that the more frequent topics were common among the majority of the sources, which is a predictable outcome. 
However, the interconnections of least frequent topics with nearly all other topics would imply that certain sources posted content on a wide array of topics. 
It may be possible to infer further that these sources in question were in the control of individuals who were more active on social media.    

\begin{figure*}
\includegraphics[trim={0 5cm 0 5cm},width=0.5\textwidth]{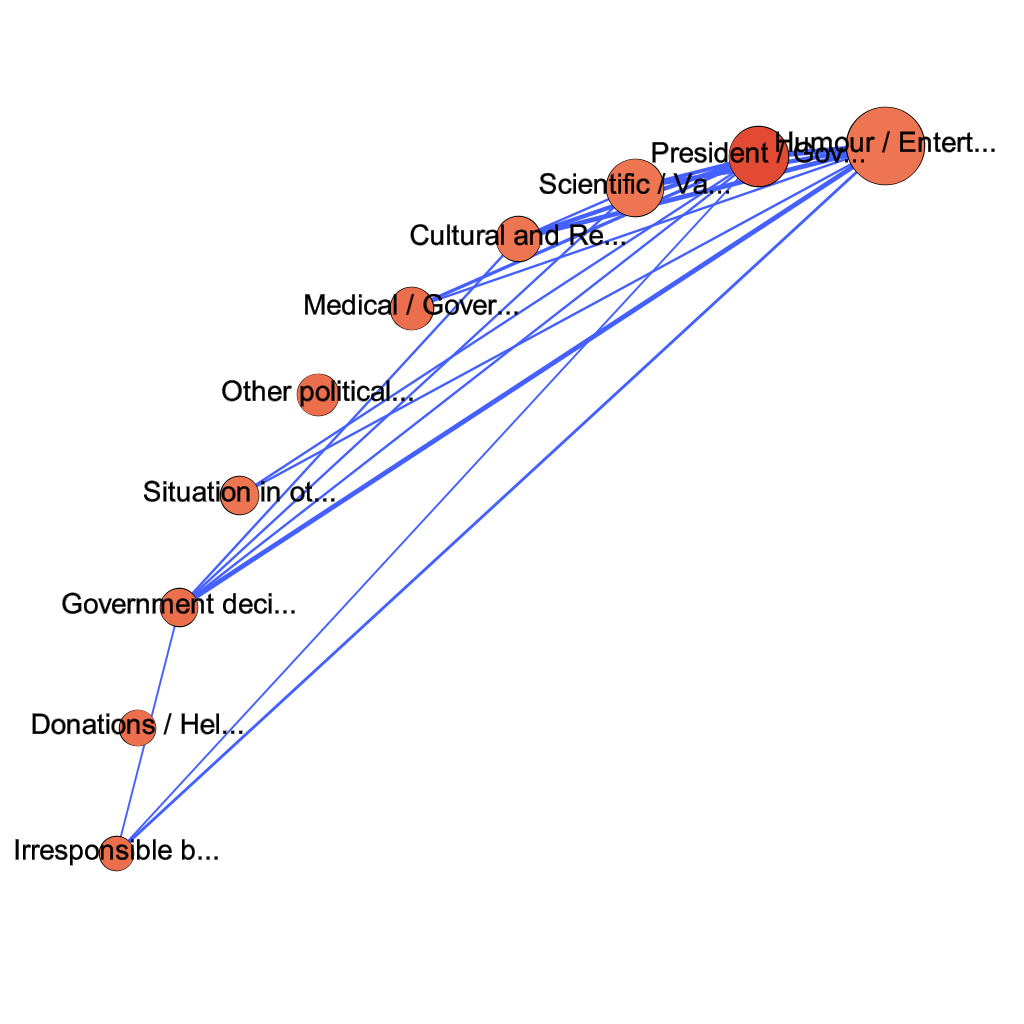}
\centering
\caption{Co-occurrence graph of the topic categories based on data sources considering co-occurrences with more than 23 sources}
\label{fig:46coo}
\end{figure*}

Another interesting point that could be observed as shown in Fig. \ref{fig:46coo}, was that when the edges were filtered to contain only weights of 23 and above, i.e. having a co-occurrence between 23 sources and above, it can be observed that topics such as \textit{Other political parties / Politicians} and \textit{Donations / Helping others / Needy} which have a higher normalised post count, seem to have a lesser co-occurrence than certain topics such as \textit{Irresponsible behaviour} that have a lower normalised post count.
This would imply that certain individuals or groups are more focused on posting content related to politics compared to the majority of the sources considered. Thus, this brings to light that certain individuals would even use a world-pandemic situation to further their own political agenda.

\subsection{Support shown for Topics}
To get an understanding of how the overall populace reacted to the various topics, an analysis was done on sources that that either supported or criticised in posts relevant to the said topics.
This was done by selecting a topic and then counting the number of negative or positive reactions among the sources. Afterwards, a score was devised for each topic, where if a source supported the topics the score rose by 1, if a source criticised a topic the score fell by 1, or else the score remained the same. This is presented by \eqref{eq:freq}, where \textit{a$_{k}^{i}$} depicts the score of the \textit{i\textsuperscript{th}} person for topic \textit{k} and \textit{A$_{k}$} depicts the total score for the \textit{k\textsuperscript{th}} topic. 
The overall results of this analysis can be seen in Fig. \ref{fig:svst} where the topics were arranged based on the score they received.

\begin{equation}
  A_{k}=\sum_{i=1}^{50}a_{k}^{i}= \begin{cases}
    +1 & \text{Supported}.\\
    -1 & \text{Criticised}.\\
    0 & \text{Neither / Both}.\\
  \end{cases}
  \label{eq:sup}
\end{equation}

\begin{figure*}
\includegraphics[trim={0 3cm 0 3cm}, width=\textwidth]{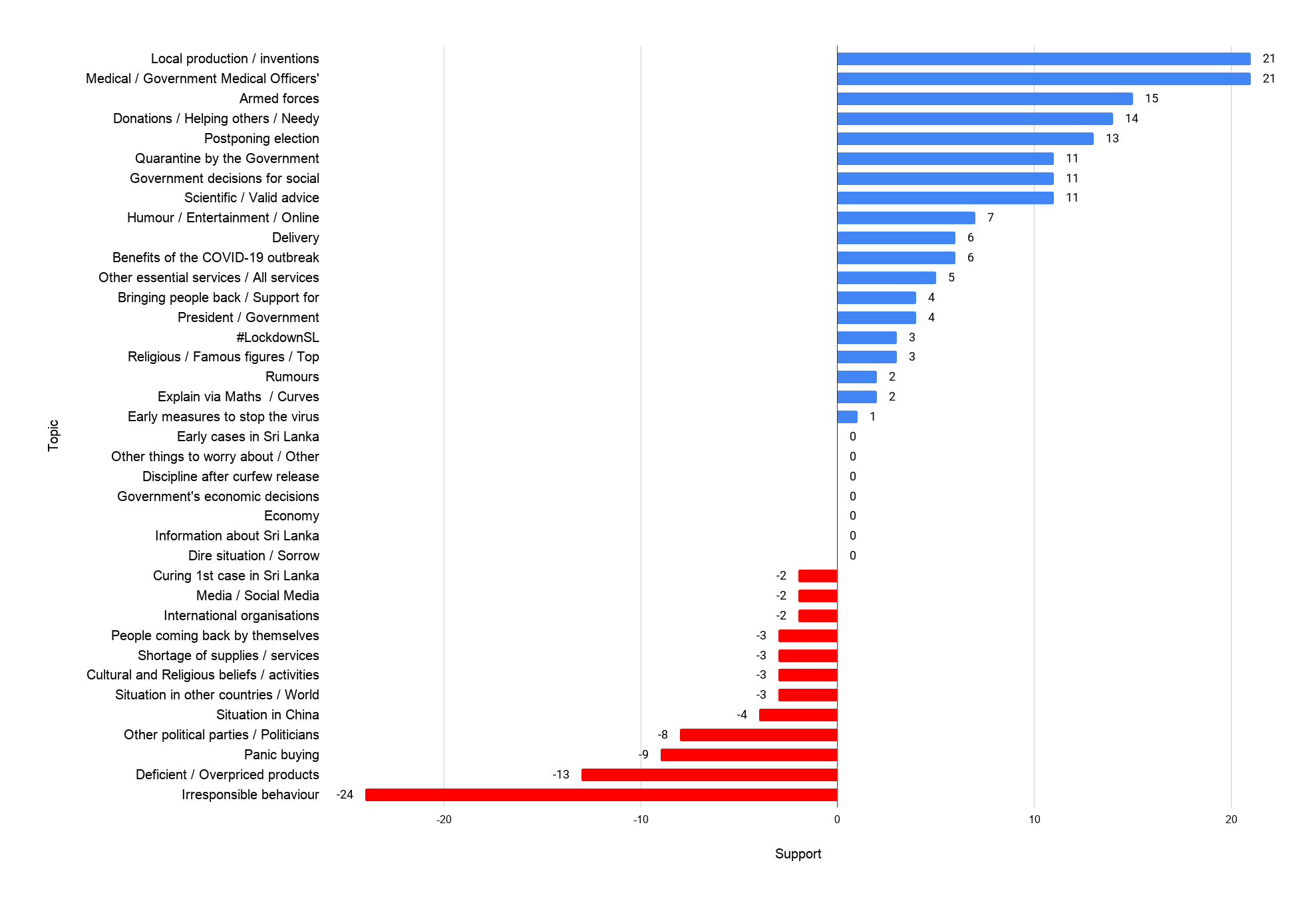}
\caption{The overall support for each topic category}
\label{fig:svst}
\end{figure*}

It could be observed that the most support in Facebook was for local productions as well as those professions that played an active role to mitigate the effects of the COVID-19 virus outbreak in Sri Lanka. Namely, these professions were that of the medical staff as well as the armed forces.  
Although showing support to professions that were the vanguard against COVID-19 outbreak was a common trend even in countries other than Sri Lanka, this usually referred to the medical staff and other essential service providers. 
Sri Lanka stood out due to the fact that mobilised its armed forces to play an active role in this crisis. Furthermore, coupled with the other most supported topic which was local production, this brings to light the nationalistic views of the people.

Following such topics were posts related to \textit{Donations / Helping others / Needy}. The support for such actions comes at no surprise. However, following even that in relevance to the support score, were certain topics related to politics such as postponing the election and government actions to curb the COVID-19 outbreak. However, these two topics were in opposition to one another. 
Due to the power shift after the 2019 presidential election in Sri Lanka, it could be said that the current government was against \textit{Postponing election}. 
However, it was supported publicly.
As the other decisions by the Government to control the spread of the virus are supported as well, the reason for the statement regarding the contradiction would be lucid if taking all the actions of the government as a whole. 
However, there were only 5 sources that truly supported all posts relating to both of these topics. 
Thus, these sources may have seen the efforts of the government in a positive light while also having a mindset of stopping any situations that were counterproductive to the actions taken thus far to curb the virus. 
However, the remaining sources may have posted content related to such topics while being under some political influence.  

Looking at the opposite end of the spectrum, the most criticised topic out of the sources considered was \textit{Irresponsible behaviour}. Such a negative response was likely due to certain actions by certain subsets of the population that completely undid all efforts of the country such as evading quarantine as well as crowding. Following this topic was \textit{Deficient / Overpriced products}, which sparked such a negative response due to the apparent unethical behaviour of certain companies utilising this crisis to rake in a large profit. The remaining topics either had a mixture of reactions or the content posted were primarily informative rather than showing any support or criticism.

\section{Discussion}

Although the manual data collection process severely limited the number of sources that were considered for the research, a notable advantage was that image posts as well as posts that indirectly related to the COVID-19 outbreak could also be identified. 
Assuming the standard automated data collection process is by a search via a query, such posts would be missed, especially when there are no clearly discernible keywords to identify them. 
The importance of this point is clearly observable when considering the results of the analysis regarding topic significance. 
To elaborate, the topic \textit{Humour / Entertainment / Online activities / Boredom relief} possessed the greatest significance out of all the topics. 
However, a considerable number of posts related to this topic such as \textit{memes} or those having no specific keywords to identify them would have otherwise been missed if not for the manual collection process. 
Thus, the value of the methodology that was used to collect data is clear, as if such posts were missed, the actual topic with the greatest significance would not have earned its true rank.

From the derived results, it was possible to observe a variety of behaviours in the sources considered. These behaviours were not only positive but also negative. Examples for negative behaviours were political polarisation, as well as unnecessary outrage and overreactions to certain incidents.
However, curtailing such detrimental behaviours as well as imposing proper regulations should be the task of the government and other relevant authorities. 
It would also be possible to utilise the general public opinions regarding government policies and preparations to overcome the COVID-19 threat that were observed in this analysis to support such regulatory measures as well.
To a certain extent, numerous issues and difficulties that people faced during the disaster situation were also identified during the analysis. 
This also gives some opinions on general circumstances that could occur in an any given disaster situation in a developing country. In addition, observations have shown that Facebook has also been used as a method of sharing valuable information as well as advice to raise public awareness and mitigate the pandemic.

A notable limitation of this research would be its limited amount of sources. Although it could be called sufficient for an exploratory study, the fact that it would not truly capture the actual behaviour of the total user-base in Facebook cannot be denied. As this method was employed due to the limitations of extracting data from Facebook, other social media platforms where it would possible to extract data by queries may offer means to perform a more rigorous study. In addition, it should also be possible to combine multiple of such sources to create a composite data-set which could be used to properly represent the Sri Lankan populace. Although granted that utilising such methods would lead to ignoring image-based social media messages, which is an advantage of the current study. However, a sufficiently large data-set possessing text-based content would also allow the usage of Natural Language Processing (NLP) techniques to perform a much more informative analysis. In addition, another major limitation of this research would be the constant fluctuation of Facebook content as the number of reactions, comments and shares could change by the second. 
Although to mitigate such issues the data extraction process was initiated approximately a week after the latter-most date considered for creating content, it does not grantee that the data sources would not change at a later date. 
Furthermore, it is also possible to delete or hide posts. This could potentially lead to considerably more severe conflicts between what was recorded and what is currently observable. 

\section{Conclusion}
This article discusses the reaction of the Sri Lankan Facebook user-base to the COVID 19 outbreak. Sri Lanka, which is a developing country in South-Asia, has approximately 70\% Facebook users out of its social media users.
Thus, it can be claimed that this study could represent the majority of the general populace on the island.
The study was conducted using over 2300 posts from 50 sources on Facebook, to observe the behaviours that were highlighted during the disaster situation. The data collected was then categorised and subsequently normalised to be subjected to analysis.
Areas such as behavioural patterns, significant topics, as well as the co-occurrence of topics were such analysis performed to generate notable insights.
Ultimately the insights presented by this study could be utilised for reinforcing effective and pragmatic decision making by governments and other relevant authorities. Furthermore, it can also be used to foster regulations and recommendations in information sharing in the context of social media.

\bibliography{IEEEpaper}

\balance

\end{document}